\documentclass[twocolumn,prb,showpacs,preprintnumbers,groupedaddress]{revtex4}%
\usepackage{mathptm}
\usepackage{dcolumn}
\usepackage{times}
\usepackage{amsfonts}%
\usepackage{amssymb}
\usepackage{bm}
\usepackage{graphicx}
\usepackage{amsmath}
\usepackage{amsfonts}%
\providecommand{\U}[1]{\protect\rule{.1in}{.1in}}
\newcommand {\be}{\begin{equation}}
\newcommand {\ee}{\end{equation}}
\newcommand {\bea}{\begin{eqnarray}}
\newcommand {\eea}{\end{eqnarray}}

\begin{document}
\title{Coexistence of superconductivity and a spin density wave in pnictides: Gap symmetry and nodal lines}
\author{D.~Parker$^{1}$}
\author{M.G. Vavilov$^{2}$}
\author{A. V. Chubukov$^{2}$}
\author{I.I.~Mazin$^{1}$}
\affiliation{$^{1}$Naval Research Laboratory, 4555 Overlook Ave. SW, Washington, DC 20375}
\affiliation{$^{2}$Department of Physics, University of Wisconsin, Madison, Wisconsin
53706, USA}
\date{\today}

\begin{abstract}
We investigate the effect of a spin-density wave (SDW) on $s_{\pm}$
superconductivity in Fe-based superconductors. We show that, contrary to
the common wisdom, no nodes open at the new, reconnected Fermi surfaces when the hole
and electron pockets fold down in the SDW state, despite the fact that
the $s_{\pm}$ gap changes sign between the two pockets. Instead, the order
parameter preserves its sign along the newly formed Fermi surfaces.   The familiar
experimental signatures of an $s_{\pm}$ symmetry are still preserved, although
they appear in a mathematically different way. For a regular $s$ case
($s_{++})$ the nodes do appear in the SDW state. This distinction suggests a
novel simple way to experimentally separate an $s_{\pm}$ state from a regular
$s$ in the pnictides. We argue that recently published thermal
conductivity data in the coexisting state are consistent with the $s_{\pm},$
but not the $s_{++}$ state.

\end{abstract}

\maketitle

The superconducting pnictides continue to attract great interest over a year
after the original discovery. Despite more than a thousand preprints and
publications, the most basic questions about pairing symmetries and mechanisms
remain controversial. Early on the s$_{\pm}$ pairing symmetry was
proposed\cite{mazin_prl,kuroki}, in which the superconducting gap function
changes sign from the hole to the electron pockets, but is roughly constant on
each surface. A possibility of an accidentally nodal s$_{\pm}$ state, or a
d-wave state, depending upon parameter values, was also proposed\cite{kuroki},
and investigated in many details recently \cite{graser}.

As of now, significant experimental evidence has been accumulated in favor of
the $s_{\pm}$ proposal, and substantial theoretical effort has been devoted to
the study of various properties of such a state (see Refs. \onlinecite{MS,C} for
reviews). So far, however, no one has addressed the possible modification of an
$s_{\pm}$ state due to a static spin density wave (SDW) coexisting with
superconductivity. At the same time the emerging consensus among
experimentalists (see Refs.
\onlinecite{PhDiag-ORNL,BK122-coex,PhDiag-Ian,Sm-PhDiag,bobroff,julien})
is that in most systems, most notably in both
electron and hole doped 122 materials, there is a range of coexistence of SDW
and superconductivity, probably up to the optimal doping level (some, however,
have argued for mesoscopic phase separation on the hole-doping side\cite{julien,phase-sep}). It was
recently estimated that the magnetic moment at the Co concentration of 7\% is
0.1 $\mu_{B}$ per Fe, corresponding, roughly, to an antiferromagnetic field of
the order of 50--100 meV~\cite{bobroff}.

The subject of an SDW coexisting with superconductivity has a long history,
dating back to Bulaevskii et al in 1980 \cite{bulaevskii} and numerous work
since then. It was shown \cite{bulaevskii,overhauser} that in a one-band BCS
superconductor a spiral SDW induces a gap anisotropy that leads to gap nodes,
while a collinear SDW still leads to a finite energy gap \cite{overhauser}.
The case of the $s_{\pm}$ superconductivity in Fe-based superconductors (FBS),
on the first glance, seems
quite simple:  first, the SDW wave we are dealing with here is simply a
collinear double-cell antiferromagnetic (AF) order, so one need not be
bothered by the difference between a spiral and collinear SDW; second,
the doubling of the unit cell in real space leads to the folding down of the
Brillouin zone in momentum space, which projects the electron Fermi
surfaces (FS) with the negative order parameter ($\Delta_{e}<0)$ onto the hole
Fermi surface with the positive order parameter ($\Delta_{h}>0)$. Whenever the
two FSs intersect,
an SDW gap opens up. It  seems obvious
that,  when that happens, $\Delta$ on the newly formed FSs should change sign,
that is, develop nodes.

However, not everything that seems obvious is true. We will show below
that, instead,  a curious novel state is formed, which is fully gapped and,
formally, has an order parameter (OP) of the same sign everywhere. This should
not be confused though with the conventional BCS-like $s$ state: when the SDW
amplitude is vanishingly small, this state \textit{has the same observable
properties as the original } $s_{\pm}$ \textit{state, despite having a
single-sign OP. }This bizarre property, which,  incidentally, is also
relevant to the coexistence of d-wave superconductivity and AF order in
electron-doped cuprates\cite{cup}, can be traced down to two facts, well known
but often not appreciated: (1) not only the overall sign of the OP in a
superconductor, but also the \textit{relative} sign of $\Delta_{\mathbf{k}}$
and $\Delta_{\mathbf{k}^{\prime}}$ is not uniquely defined, but depends on the
convention for the wave function phases and (2)  as opposed to a nonmagnetic
material, in an AF metal it is not possible to fix the phases of the wave
functions in such a way that the wave functions for both spin projections are
identical at any \textbf{k}-point.

With these considerations in mind, let us now outline the derivation. We will
follow the approach of Ref. \onlinecite{overhauser}, and for illustrative purposes
will use a simple semimetallic model bandstructure with a hole band centered
at the $\Gamma$
point and an electron band around
$\mathbf{Q}=$($\pi,0$) and related points in the unfolded Brillouin zone
which we will be using
. We assume the Fermi energies to be, respectively, $\epsilon_{h}$ and
$\epsilon_{e},$ and take an isotropic effective mass $m$ for the hole band. To
account for the fact that nesting is always imperfect (and if it were perfect,
the SDW would open a gap on the entire FS, thus preventing any coexisting
superconductivity), we take $m_{e}$ to be anisotropic, with $m_{x}\neq m_{y}$.
This reflects the fact that the actual calculated and measured anisotropy of
the electron pockets is larger than that of the hole ones. The Hamiltonian for
this system is given by \cite{bulaevskii}
\begin{equation}
\hat{H}=\sum_{k,i}\epsilon_{k,i}\varphi_{k,i}^{\dagger}\varphi_{k,i}%
+\mathbf{h}_{Q,i}\cdot\mathbf{S}_{k}%
\end{equation}
with an effective field $h\propto\cos(\mathbf{Qr})$, which interacts with the
electron spin $\mathbf{S}$ and leads to the SDW. The index $i=e,h$ refers to
the hole or electron Fermi surface bands, with the dispersions \vspace
{-0.1cm}
\begin{equation}%
\begin{split}
\epsilon_{k,e} &  =\hbar^{2}(k_{x}-\pi)^{2}/2m_{x}+\hbar^{2}k_{y}^{2}%
/2m_{y}-\epsilon_{e},\\
\epsilon_{k,h} &  =-\hbar^{2}k^{2}/2m+\epsilon_{h}\vspace{-0.2cm}.
\end{split}
\end{equation}
For the purpose of this work -- demonstrating the effect of AF -- we will take
$\epsilon_{h}$ and $\epsilon_{e}$ equal. In real life, of course, they will
depend on the relative location of $E_{F}$ which will change with doping. We
take $m_{y}<m<m_{y}$ to ensure intersections between the hole and electron
Fermi surfaces upon translating by the SDW vector. We take $m_{x}m_{y}\sim
m^{2}$ so that the electron and hole densities of states (DOS) are
comparable.\cite{param}

In the following we will work in the downfolded Brillouin zone corresponding
to the antiferromagnetic unit cell. Let $G$ be the matrix element (assumed to
be $k-$independent) of the SDW potential mixing the hole and electron wave
functions, $\varphi_{h}$ and $\varphi_{e}.$ Then the dispersion in the SDW
state is{%
\begin{equation}
E_{k}^{\pm}=\frac{\epsilon_{\mathbf{k},h}+\epsilon_{\mathbf{k},e}\pm
\sqrt{(\epsilon_{\mathbf{k},h}-\epsilon_{\mathbf{k},e})^{2}+4G^{2}}}{2}
\label{en}%
\end{equation}
with the } new wave functions
\begin{align}
\psi_{k\uparrow}^{+}  &  =\cos\theta_{k}\varphi_{k,h}+\sin\theta_{k}%
\varphi_{k,e};~\psi_{k\uparrow}^{-}=\sin\theta_{k}\varphi_{k,h}-\cos\theta
_{k}\varphi_{k,e}\label{psi1}\\
\psi_{k\downarrow}^{+}  &  =\cos\theta_{k}\varphi_{k,h}-\sin\theta_{k}%
\varphi_{k,e};~\psi_{k\downarrow}^{-}=\sin\theta_{k}\varphi_{k,h}+\cos
\theta_{k}\varphi_{k,e} \label{psi2}%
\end{align}
where $\tan\theta_{\mathbf{k}}=G/(E_{\mathbf{k}}^{-}-\epsilon_{\mathbf{k},e}%
)$. {In Figure 1 we plot the separate hole and electron Fermi surfaces above
the SDW ordering temperature (main panel), as well as the Fermi surface for
two values of }$G.$ Of course, {we are most interested in the limit in which
the SDW-created reconstruction of the Fermi surface is relatively minor. In
practical terms the SDW gap may be (albeit not necessary for retaining
superconductivity!) smaller than the SC gap, but the concept is easier to
illustrate for an SDW gap comparable to the superconducting one.
\begin{figure}[h]
{ \includegraphics[width=8cm]{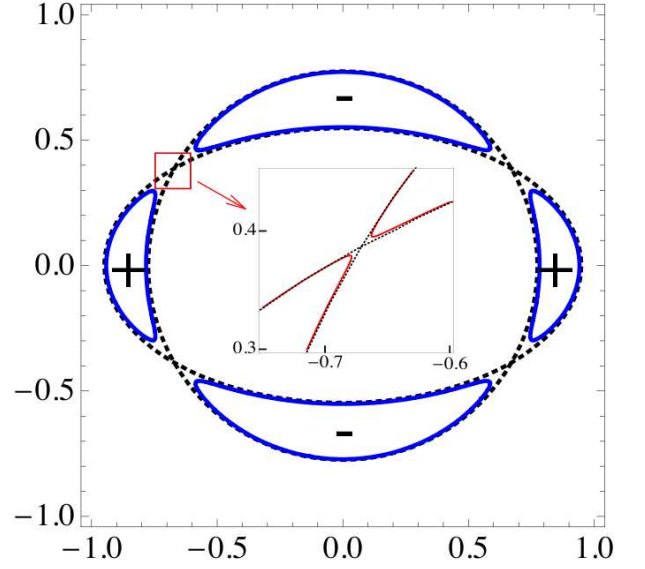} }\caption{(Color online) Main figure:
heavy line, the Fermi surface in the SDW state for $G=E_{F}$/12; dashed line,
the Fermi surface above the SDW ordering temperature.   The ``$+$" and ``$-$" Fermi surfaces are
indicated.  Inset: the Fermi
surfaces for $G=E_{F}/120$.}%
\end{figure}}Figure 2 shows the behavior of $\theta_{k}$ along the SDW-state
Fermi surfaces. Note that while the energy spectrum (\ref{en}) is double
degenerate, the wave functions (\ref{psi1}) are not equal to the the wave
functions (\ref{psi2}). If we now create a singlet anomalous average (the
superconducting OP), it will  look like
\begin{align}
&  \Delta^{+}(\mathbf{k})\mathbf{=}\left\langle \psi_{k\uparrow}^{+}%
\psi_{-k\downarrow}^{+}\right\rangle =\cos^{2}\theta_{k}\left\langle
\varphi_{k,h}\varphi_{k,h}\right\rangle \label{Dplus}\\
&  -\sin^{2}\theta_{k}\left\langle \varphi_{k,e}\varphi_{k,e}\right\rangle
\ +2\sin\theta_{k}\cos\theta_{k}\left\langle \varphi_{k,e}\varphi
_{k,h}\right\rangle ,\nonumber
\end{align}
and a similar expression for $\Delta^{-}(\mathbf{k})$. At $G\rightarrow0,$ on
the new \textquotedblleft+\textquotedblright\ FS pockets $\theta_{k}$ takes
the value of 0 or $\pi/2$ depending on whether the point on the FS originated
from holes or from electrons (and the reverse holds for the \textquotedblleft%
$-$\textquotedblright\ pockets). Recalling the original $s_{\pm}$ assumption,
$\Delta^{h}=-\Delta^{e}$ with $\Delta^{i}\mathbf{=}\left\langle \varphi
_{i}\varphi_{i}\right\rangle $,  $\left\langle \varphi_{h}\varphi
_{e}\right\rangle =0$ we immediately observe that $\Delta^{+},$ as opposed to
$\Delta^{h,e},$ never changes sign!

However, the system remembers all too well that part of the new FS has come
from electrons and part from holes, and that these used to have the OPs of the
opposite sign. If we try to calculate any observable quantity (the OP $per$
$se$ is not observable), such as pair scattering from one part of the FS to
another, we will have to take into account the fact that the new wave
functions for the up spin (\ref{psi1}) are approximately equal to the old wave
functions (e- or h-, depending on what part of the pocket we consider), but
for the down spin (\ref{psi2}) the same holds for the hole-type part of the
FS, while for the electron-type parts the sign of the wave function is
flipped. That is to say, any observable matrix elements include a product of
the OPs \emph{and} of the one-particle wave function; in the original,
unfolded bands, we were able to choose signs of the
wave functions
to be spin-invariant, so that the scattering from the hole to the electron FS
involved a sign change ($s_{\pm}).$ However, in the new, AF zone, no sign change is
generated by the OPs, but the same sign change necessarily appears from the
normal part of the matrix elements.

We see that we cannot describe the system even with infinitesimally weak SDW
in terms of the same wave functions we used for the nonmagnetic parent system.
Yet one can restore the conceptual continuity by describing the $parent$
system differently. Let us select the wave function phases in the nonmagnetic
system so that $\varphi_{k,h\downarrow}(\mathbf{r})=\varphi_{k,h,\uparrow
}(\mathbf{r}),$ but $\varphi_{k,e,\downarrow}(\mathbf{r})=-\varphi
_{k,e,\uparrow}(\mathbf{r}).$ This is a very inconvenient, but legitimate
gauge. In this gauge, the OP in the $s_{\pm}$ state will have the same sign on
the both FSs, but any physical observable involving pair scattering will have
to account for a sign flip for electrons, but not holes, and thus the overall
result will be unchanged --- the same physical situation that we find (and in
that case cannot avoid) in the SDW state.

Let us move on to an arbitrary strength SDW and evaluate \cite{footnote} the
pairing matrix involved in a superconducting state below the SDW ordering
temperature, $\Lambda_{kk^{\prime}}^{\alpha\beta}=-\langle\psi_{k\uparrow
}^{\alpha}|U|\psi_{k^{\prime}\uparrow}^{\beta}\rangle\langle\psi_{k^{\prime
}\downarrow}^{\beta}|U|\psi_{k,\downarrow}^{\alpha}\rangle,$ where $U$ is the
pairing interaction, with all relevant factors included and $\alpha,\beta
=\pm1$ are new band indices. We assume that only hole-electron matrix elements
$\langle\varphi_{k,h}|U|\varphi_{k^{\prime},e}\rangle$ are nonzero, and the
minus in front accounts for the fact that the pairing interaction is assumed
to be generated by spin fluctuations. We furthermore assume that this matrix
element does not depend on $k,$ $k^{\prime}$.\cite{RG}

After some trigonometric manipulations, we get the answer:%
\begin{equation}
\Lambda_{kk^{\prime}}^{\alpha\beta}=\frac{V}{2}(1-\alpha\beta\cos2\theta
_{k}\cos2\theta_{k^{\prime}}+\alpha\beta\sin2\theta_{k}\sin2\theta_{k^{\prime
}})\nonumber
\end{equation}
The factor 1/2 was selected so that (as we will show later) the effective
coupling constant in the $G=0$ limit will be equal to $V\sqrt{N_{e}N_{h}}.$ Now we
can write the BCS equation at $T=T_{c}$ as (with $\omega$ being the BCS cutoff
energy in temperature units)%

\begin{align}
&  {\ln(1.13\omega/T_{c})}\Delta_{k}^{\alpha}=\frac{V}{2}\sum_{k^{\prime}%
\beta}\delta(E_{k^{\prime}}^{\beta})\Delta_{k_{\beta}^{\prime}}\label{1}\\
&  -\frac{V}{2}\alpha\cos2\theta_{k}\sum_{k^{\prime}\beta}\beta\delta
(E_{k^{\prime}}^{\beta})\cos2\theta_{k^{\prime}}\Delta_{k_{\beta}^{\prime}%
}\nonumber\\
&  +\frac{V}{2}\alpha\sin2\theta_{k}\sum_{k^{\prime}\beta}\beta\delta
(E_{k^{\prime}}^{\beta})\sin2\theta_{k^{\prime}}\Delta_{k^{\prime}}^{\beta}%
\end{align}
We
seek the solution of this equation in the following form
\begin{equation}
\Delta_{k}^{\alpha}/\Delta_{0}=c+\alpha a\cos2\theta_{k}+\alpha b\sin
2\theta_{k}, \label{main}%
\end{equation}
and
\begin{align}
{\ln(1.13\omega/T_{c})}c  &  =\frac{V}{2}(aN_{c}+bN_{s}+cN)\nonumber\\
{\ln(1.13\omega/T_{c})}a  &  =-\frac{V}{2}(aN_{c2}+bN_{s}+cN_{c})\label{sec}\\
{\ln(1.13\omega/T_{c})}b  &  =\frac{V}{2}(aN_{cs}+bN_{s2}+cN_{s}\}\nonumber
\end{align}
{where we introduced the weighted DOSs: }$N=\sum\delta(E_{k}^{\alpha}),$
$N_{c}=\sum\alpha\delta(E_{k}^{\alpha})\cos2\theta_{k},$ $N_{s}=\sum
\alpha\delta(E_{k}^{\alpha})\sin2\theta_{k},$ $N_{c2}=\sum\delta(E_{k}%
^{\alpha})\cos^{2}2\theta_{k},$ $N_{s2}=N-N_{c2}=\sum\delta(E_{k}^{\alpha
})\sin^{2}2\theta_{k},$ $N_{cs}=\sum\delta(E_{k}^{\alpha})\cos2\theta_{k}%
\sin2\theta_{k}.$ All summations are over $k$ and $\alpha$, and $\delta$
represents the Dirac delta function.

The maximal eigenvalue of the matrix
\begin{equation}
\lambda=\frac{V}{2}\left(
\begin{array}
[c]{ccc}%
N & N_{c} & N_{s}\\
-N_{c} & -N_{c2} & -N_{cs}\\
N_{s} & N_{cs} & N_{s2}%
\end{array}
\right)  , \label{l}%
\end{equation}
as usual, defines the BCS transition temperature $T_{c},$ and the
corresponding eigenvector gives the distribution of the order parameter over
the Fermi surfaces.

In the limit $G\rightarrow0,$ $\cos2\theta_{k}\rightarrow\pm1$, $N_{c}%
\rightarrow N_{e}-N_{h},$ $N_{c2}\rightarrow N,$ and all other DOSs vanish.
Here $N_{e(h)}$is the density of states on the electron (hole) Fermi surface
without the SDW. The effective coupling constant $\lambda_{eff}=V\sqrt
{N^{2}-N_{c}^{2}}/2=V\sqrt{N_{e}N_{h}},$ which is the well-known result for
the $s_{\pm}$ pairing.  The ratio of the two gaps at $G\rightarrow0$ is
$\sqrt{N_{h}/N_{e}}$ as it should be in a weak-coupling $s_{\pm}$
superconductor\cite{gaps}.

At a finite $G$, the gap does become angle-dependent, via $\theta_{k}$, but it
is easy to prove\cite{prove} that the eigenvector for maximal $\lambda$ in
(\ref{1}) gives $c>a\gg b,$ so that the gap  has the same sign everywhere, in
agreement with general discussion after Eq.~(\ref{Dplus}).  The gap is depicted in Figure 2, where we note that
\begin{figure}[h]
\includegraphics[width=8cm]{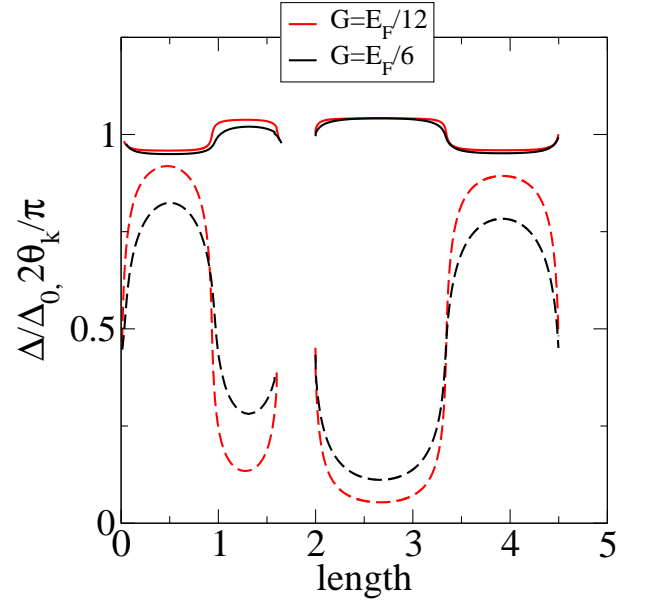}
\caption{(Color online) Depicted is the variation of $\Delta({\bf k})$ (solid lines) and $\theta_{k}$
(dashed lines) along the Fermi surfaces for the indicated values of $G$.  Note that the angle $\theta_{k}$ will approach a step function, with regions where $\theta_{k}=0$ or $\theta_{k}=\pi/2$, as $G \rightarrow 0$.  Here ``length" refers to the k-space arc-length around the Fermi surface; note that the actual arc-length depends on G.  To facilitate comparison we have scaled the arclengths for these two cases lengths to be equal; actual $G=E_{F}/6$ arclengths are $\sim 20$ percent smaller than for $G=E_{F}/12$.}%
\end{figure}
the gap changes rapidly within a finite area where the SDW gap opens; the length of this region
scales with G.

We have also found that
the effective coupling constant is reduced slightly at a finite $G$ compared to its
value at $G=0$, i.e., superconducting $T_{c}$ drops in the presence of an SDW
order. At small $G$, we found $\delta\lambda_{eff}\approx-(VN_{s2}%
/8)(\sqrt{N_{e}}-\sqrt{N_{h}})^{2}/\sqrt{N_{e}N_{h}}$ and\cite{prove}
$N_{s2}\propto|G|/|N_{e}-N_{h}|$. Note that the dependence on $G$ is non-analytic.

The calculations above can be easily generalized to three other cases: an
$s_{\pm}$ with a charge density wave (CDW), a regular $s$ with a SDW, and a
regular $s$ with a CDW. The last case is mathematically equivalent to the one
considered above. The first two cases are equivalent to each other and
constitute a different set. Indeed, in that case the $V$ in Eqs. \ref{1},
\ref{sec}, \ref{l} should have the opposite sign, and the solution will always
have eight node lines at the tips of the banana-shaped FSs of Fig. 1. The
eigenvector for the largest eigenvalue will have, for $N_{e}\sim N_{e},$ the
largest weight on $a,$ not $c,$ and the OP will have one sign for the parts of
the FS that originated from the electrons and the opposite sign for those
originating from the holes.

To summarize, we have shown that, surprisingly, the SDW observed in underdoped
pnictide compounds, does not have any considerable destructive effect on the $s_{\pm}$
superconductivity, besides the obvious competition between the two
instabilities for the density of states at the Fermi level. As opposed to a
hypothetical CDW, which would have created
nodes on the FS and additionally weakened superconductivity, an SDW wave retains
the gapped nature of the $s_{\pm}$ superconductivity. Nevertheless, this
constant-sign state has the same observable physical properties as the
sign-changing $s_{\pm}$ state without an SDW.
In particular, the penetration depth $\lambda(T)\propto1/\sqrt{\rho_{s}(T)}$
is still exponential at the lowest $T$ and crosses over to roughly $T^{2}$ at
higher $T$~\cite{vorontsov}. The slope of $\delta\lambda(T)\propto\delta\rho
_{s}(T)/(\rho_{s}(T=0))^{3/2}$ however increases with increasing SDW order
simply because $\rho_{s}(T=0)$ decreases together with the area of the Fermi
surface.   This slope increase upon approaching the SDW state was observed
in Ref. \onlinecite{gordon2}.  If, however, the pairing state
were a conventional $s$-wave, an SDW order
would give rise to gap nodes, and $\delta\lambda$ would be linear in $T$ at
the smallest $T$ (we have explicitly verified that the SDW coherence factors
do not affect this behavior). Similarly, the reduced thermal conductivity
$\kappa/T$ should still vanish at $T\rightarrow0$ if the gap is $s_{\pm}$ but
should become finite below the onset of the SDW order if the gap is a
conventional $s$-wave, and the more we go into the underdoped regime, the
larger the residual $\kappa/T$ should be because its value does not depend on
impurity concentration, but only on the inverse of the slope of $\Delta_{\mathbf{k}}$ near
the nodes.
Indeed, a recent study of the low-$T$ thermal conductivity
in the underdoped BaFe$_{2-x}$Co$_{x}$As$_{2}$, where microscopic coexistence
of SDW and superconductivity has been well documented, shows~\cite{kappa}
that
$\kappa/T$ vanishes at $T\rightarrow0$ indicating absence of any gap nodes.
This is consistent, according to our results, with an $s_{\pm}$ state.

We acknowledge useful discussions with P. Hirschfeld, I.Eremin, and D.
Scalapino. The work was supported by NSF-DMR 0604406 (A.V.C) and by the Office
of Naval Research.





\begin{thebibliography}{99}   

\bibitem {mazin_prl}{ {I.I. Mazin, D.J. Singh, M.D. Johannes, and M.H. Du, Phys. Rev. Lett.
\textbf{101}, 057003 (2008).}}
%


\bibitem {kuroki}K. Kuroki, S. Onari, R. Arita, H. Usui, Y. Tanaka, H. Kontani,
and H. Aoki, Phys. Rev. Lett. \textbf{101},
087004 (2008).
%


\bibitem {graser}{S. Graser, T. A. Maier, P.J. Hirschfeld and D.J.
Scalapino,  New J. Phys. \textbf{11}, 025016
(2009);  A.V. Chubukov, M.G. Vavilov and A.B. Vorontsov, arXiv:0903.5547.}

\bibitem {MS}I.I. Mazin and J. Schmalian, Physica C, 469, 614 (2009)



\bibitem {C}A.V Chubukov, D. Efremov and I. Eremin, \prb, \textbf{78}, 134512
(2008); V. Stanev, J. Kang, and Z. Tesanovic, Phys. Rev. B {\bf 78}, 184509 (2008);
A.V. Chubukov, Physica C, {\bf 469}, 640 (2009).


\bibitem {PhDiag-Ian}{ J.-H. Chu, J.G. Analytis, C. Kucharczyk, and I. R.Fisher, {Phys. Rev.} B \textbf{79},
014506 (2009). }


\bibitem {Sm-PhDiag}{A. J. Drew, Ch. Niedermayer, P. J. Baker, F. L. Pratt,
S. J. Blundell, T. Lancaster, R. H. Liu, G. Wu, X. H. Chen, I. Watanabe,
V. K. Malik, A. Dubroka, M. R̦ssle, K. W. Kim, C. Baines and C. Bernhard, Nature Materials \textbf{8,
}310 (2009);
S. Sanna, R. De Renzi, G. Lamura, C. Ferdeghini, A. Palenzona,
M. Putti, M. Tropeano and T. Shiroka,  arXiv:0902.2156 (unpublished);
S. Takeshita, R. Kadono, M. Hiraishi, M. Miyazaki, A. Koda, Y. Kamihara,
and H. Hosono,  J. Phys. Soc. Jpn. \textbf{77, }103703 (2008). }

\bibitem {BK122-coex}{ M. Rotter M. Tegel, I. Schellenberg,
F.M Schappacher, R. P\"ottgen, J. Deisenhofer,
A. G\"unther, F. Schrettle, A. Loidl,
and D. Johrendt, New J. Phys. \textbf{11,}
025014 (2009). }
%

\bibitem {PhDiag-ORNL}F. Ning, K. Ahilan, T. Imai, A.S. Sefat, R. Jin, M.A. McGuire, B.C. Sales,
and D. Mandrus, J. Phys. Soc. Jpn. \textbf{78},
013711 (2009).
%
\bibitem {bobroff}Y. Laplace, J. Bobroff, F. Rullier-Albenque, D. Colson, and A. Forget, arXiv:0906.2125.

\bibitem{julien} M.-H. Julien, H. Mayaffre, M. Horvatic, C. Berthier, X.D. Zhang, W. Wu, G.F. Chen, 
N.L. Wang and J.L. Luo, arXiv:0906.3708.
%
\bibitem {phase-sep}{ J. T. Park, D. S. Inosov, Ch. Niedermayer, G. L. Sun, D. Haug, N. B. Christensen, R. Dinnebier, A. V. Boris, A. J. Drew, L. Schulz, T. Shapoval, U. Wolff, V. Neu, X. Yang, C. T. Lin, B. Keimer, and V. Hinkov,
Phys. Rev. Lett.
\textbf{102}, 117006 (2009). }



%

\bibitem {bulaevskii}L.N. Bulaevskii, A.I. Rusinov and M. Kuli\'{c}, J. Low
Temp. Phys. \textbf{39}, 255 (1980), also Sol. State Comm. \textbf{30}, 59
(1979); M.L. Kuli\'{c}, A.I. Lichtenstein, E. Goreatchkovski,
and M. Mehring, Physica C \textbf{244}, 185 (1995);
%
\textit{ibid}, \textbf{252}, 27 (1995).

\bibitem {overhauser}{ {L.L. Daemen and A. W. Overhauser, Phys. Rev. B
\textbf{39}, 6431 (1989). } }




\bibitem {cup}This case is studied by J.-P. Ismer, I. Eremin, E. Rossi, D. Morr and G. Blumberg, arXiv:0907.1296.

\bibitem {param}In the calculation shown in this paper, the parameters are:
$\varepsilon_{p}=\varepsilon_{e}=0.6,$ $m_{x}=5m/3,$ $m_{y}=m/2,$ and the bare
order parameter
$|\Delta_{0}|=0.03$.

\bibitem {footnote}We note that the approach we follow here, in which we pair
the electrons from SDW eigenstates, suggests that $T_{SDW}\gg T_{c}$, or more
precisely, that $G\gg\Delta(\mathbf{k})$. In Ref. \onlinecite{bulaevskii} the
problem of explicit coexistence of SDW and SC was considered, which results in
a minor change to the dispersion relation of order $A\frac{\Delta
(\mathbf{k})^{2}G^{2}}{E_{F}^{2}(\Delta(\mathbf{k})^{2}+G^{2})}$ in the
coexistence state, where A is a constant factor dependent on the mass
anisotropy. This complicates the derivations but does not alter the
qualitative conclusions. Given that our model is oversimplified compared to
the actual multiband material, we feel it sufficient to present an analysis in the
$G\gg\Delta(\mathbf{k})$ regime.



\bibitem {RG}This is equivalent to the assumption that the interaction depends
only on the transvered momentum but not on the total incoming momentum. In
``g-ology'' models~\cite{C} this implies that the pair-hopping and the
backscattering interaction are taken equal. We verified that the conclusions
do not change if the two interactions split under RG flow. The only difference
in this situation is a reduction of the $\sin2 \theta_{k}$ term in Eq.
\ref{main}.

\bibitem{gaps}O.V. Dolgov, I.I. Mazin, D. Parker and A.A. Golubov, Phys. Rev. B \textbf{79}, 060502
(2009).
%


















\bibitem {prove}One can show that in the limit of small $G,$ the parameters of
the coupling matrix behave as: $N_{s2}/N\approx|G|\cos^{-1}(\eta)/\pi
E_{F}\eta$ where $\eta$ is the relative variation of the electron mass, and
$N_{cs}/N\approx|G|[\eta-\sin^{-1}(\eta)]/\pi E_{F}\eta^{2}.$ For small
electron-hole anisotropy, $\eta\rightarrow0$ and $N_{cs}/G\rightarrow0$
linearly in $\eta.$ The other DOSs involve a summation over sign-changing
functions and are even smaller.

\bibitem {vorontsov} A.B. Vorontsov, M.G. Vavilov and A.V. Chubukov, Phys. Rev. B {\bf 79}, 140507(R) (2009).



%
\bibitem {gordon2}{ {R.T. Gordon, C. Martin, H. Kim, N. Ni, M. A. Tanatar, J. Schmalian,
I. I. Mazin, S. L. Bud'ko, P. C. Canfield, and R. Prozorov, Phys. Rev. B \textbf{79},
100506 (2009). } }


\bibitem {kappa}M. A. Tanatar, J.-Ph. Reid, H. Shakeripour, X. G. Luo, N.
Doiron-Leyraud, N. Ni, S. L. Bud'ko,  P. C. Canfield,  R.
Prozorov, and L. Taillefer, arXiv:0907.1276

\end{thebibliography}
\end{document}